\documentclass[twocolumn,showpacs,preprintnumbers,amsmath,amssymb,pre]{revtex4}
\usepackage{dcolumn}
\usepackage{bm}
\usepackage{epsfig}

\begin{document}
\preprint{APS/PRE}
\title{Long time viscosity of dilute magnetorheological dispersions under periodic magnetic perturbations}

\author{F. Donado}
\affiliation{Instituto de Ciencias B\'asicas e Ingenier\'{\i}a-CIMA  \\  Universidad Aut\'onoma del Estado de Hidalgo \\
Pachuca 42090, Pachuca, M\'exico}
 
\author{J.L. Carrillo}
\affiliation{Instituto de F\'{\i}sica, Universidad Aut\'onoma de Puebla
A.P. J-48, Puebla 72570, Puebla, M\'exico}


\begin{abstract}
The effect of periodic magnetic perturbations on the rheological properties of a low concentration magnetorheological dispersion is studied experimentally. It is found that an important increment in the measured viscosity occurs when in addition to a static field a magnetic periodic perturbation is applied. The magnitude of these changes depend on the amplitude and frequency of the perturbation as well as on the simultaneity of the application of the static field and the perturbation. These findings are discussed in terms of the observed rearrangement of the cluster structure in the dispersion. 
\end{abstract}
\pacs{83.80.Gv; 45.70.Qj, 83.60.Np}
\maketitle

\section{Introduction}

Magnetorheological fluids (MRF) experiment strong changes in their physical properties, mainly in the mechanical ones, when they are exposed to external magnetic fields\cite{kitti,pre,martin3,taostrong,seval}. The change in the shear modulus of the fluid can be of several orders of magnitude, in such a way that under the influence of a static field the system transforms from a viscous liquid into almost  a solid body. These complex fluids are composed by micrometric magnetizable particles dispersed in an inert newtonian oil, preferably of low viscosity. The changes induced by the applied field  are caused by the formation of structure in the dispersion. As result of this, the viscous fluid transforms into a viscoelastic one \cite{taostrong,seval,ref21,vicente}. There exist the electric analogous systems, the electrorheological fluids\cite{tao,chen,martin2,halsey2}, these are formed by polarizable particles dispersed in a liquid whose dielectric constant contrasts with that of the particles. Both classes of systems show qualitatively analogous behavior and they have attracted the interest of applied researchers to design a wide variety of devices like dampers, clutches, etc. 
From the basic science point of view, these phenomena are relevant because the complex pattern formation which takes place in the dispersion, is still not well understood. In a wider sense, the understanding of the complex pattern formation by interacting particles, in non equilibrium conditions, where jamming like phenomena occur has been, since a long time, a fundamental subject in the physics and chemistry of non homogeneous systems. 

Upon the application of a magnetic field on a MRF, a dipolar moment is induced in the suspended particles, then, they interact and aggregate forming elongated structures oriented along the applied field direction \cite{pre,ukai,micromech,furst,silva,martin1,cutillas2}. The features of the cluster structure produced by the aggregation processes strongly depend on two quantities, the strength of the applied field and the particle concentration. 

It has been shown that in low particle concentration MRF, in the presence of a static magnetic field, the aggregation process occurs basically in two sequential steps. In the first and quickest one, the particles aggregate forming clusters that closely resembles the form of chains. In the second step, these chains aggregate laterally to form columns or larger chains \cite{taostrong,furst,micromech}. Presently it is commonly accepted that the lateral aggregation of these clusters is the dominant kinetic process that governs the physical behavior of the system at the long time stages. For moderate and high particle concentrations, the process of aggregation is more complex. In some cases it has been described as consisting of at least three steps in which clusters of different sizes and fractal characteristics are formed sequentially in the suspension \cite{pre,stat}.

On the other hand, it has been observed that when a time-dependent field is applied some other important changes in the pattern formation and the rheological properties of the system can be induced. For instance, a periodic magnetic field is able to drive the formation of structures with frequency dependent characteristics \cite{dwirtz}, and whose configuration has been discussed to have a lower energy, compared to that formed by a static field \cite{joanne}. Also it has been observed that an abrupt application of an intense field originates a relatively more complex structure than those generated by the application of a field whose amplitude increases slowly, consequently, in this latter situation, the viscosity acquires a larger value comparatively to the former situation \cite{chaker}.

Some discussion regarding the origin of the interactions which produce the lateral aggregation has been reported in the literature. According to some authors, lateral interactions could be promoted by thermal fluctuations, hydrodynamic fluctuations, and defects in the chains \cite{cutillas2,martin1,silva,furst}. Some other studies have addressed the discussion on how external agents like sudden pressure increments, can change the structure formed by the particles, producing thicker clusters and consequently, modifying the rheological properties of the MRF\cite{taostrong}. 

In this work low particle concentration MRFs are studied under the influence of a magnetic oscillatory perturbation, in addition to a static magnetic field. In the presence of both fields, one of the most remarkable results is that it is possible to influence and at some extent control  the lateral aggregation. This is so because the static field is not turned off during the whole process, then, at any stage the system is able to support a shear stress. The application of both fields is a relevant difference of this work in comparison to other studies in which only one of these fields is present.

In section II the experimental procedure followed to analyze by optical microscopy the structural characteristics of the clusters formed in the suspension is presented. On this basis, in section III the viscosity measurements are discussed. Finally, the conclusions and some comments and remarks are presented.

\section{Cluster structure under perturbations} 

In order to obtain some insight on the relation between the cluster structure and viscosity, firstly it was studied by optical microscopy, under different conditions, the structural features of the clusters that the particles build in the suspension. For the observations it was used a microscope Meiji EMZ-TR and a Diagnostic Instruments digital camera. The suspensions were prepared with magnetite particles, whose measured average size was 66 $\mu$m with a dispersion of 15 $\mu m$ dispersed in mineral oil. A rectangular open cell made by cover glass with a width of 14 mm and a length of 19 mm was prepared and the dispersion was purred in, in an amount enough to form a wet layer. Particles are massive enough to precipitate quickly, forming a bidimensional distribution. Due to the low particle concentration, 0.05 in surface fraction, with no applied field the particles remain dispersed and there was no formation of clusters. To study the changes in the structure of the dispersion, the fields were applied transversal each other, both on the horizontal plane. The oscillatory field, conceived as a perturbation, had  a strength much lower than the static one. The fields were generated by means of two couples of Helmholtz coils.  

Fig. 1 shows photographs of clusters formed under the following conditions. In (a) there appears the initial dispersion of the particles in absence of applied fields. In photograph (b) it is shown the typical clusters chain-shaped for low particle concentration. The chains were formed due to the application of a  $80 G$ static field. Photograph (c) shows the effect of a transversal perturbation, it was an oscillatory $12 G$ RMS magnetic field, with a $4 Hz$ frequency, which was applied some time after, and in addition, to a $80 G$ static field. In order to contrast these effects, in (d) photo it is shown the effect of a more intense static field, $100 G$, with no  perturbation. As expected, the length of the chains clearly increases with the intensity of the applied field, it can be observed by comparing the (b) and (d) pictures. On the other hand, from the comparison between (c) and (d) one may conclude that the presence of the perturbation favors the formation of larger chains, even larger than those that would be obtained by stronger static fields. The comparison between (d), where the field is stronger than the total magnitude of the fields in (c), exhibits the important role of the perturbation in the pattern formation. This point shall be discussed below. In all of these cases the photographs were obtained at a time, after the application of the fields, when no further noticeable changes were observed.    

\begin{figure}[!tbp]
\centering
\begin{center}
\leavevmode 
\psfig{file=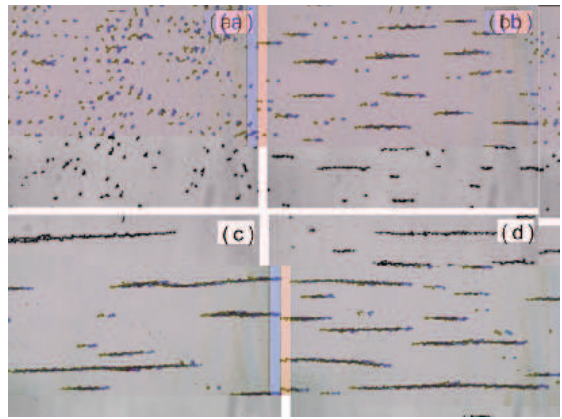,width=7.4cm} \caption{\small Clusters formed under different conditions. a) No applied field. b) $80 G$ static field. c)  $80 G$ static field and $4 Hz$, $12 G$ oscillatory perturbation applied with some delay. d) $100 G$ static field with no perturbation. } \label{chains}
\end{center}
\end{figure}

In the short time scale after the field is turned on, it is observed that the system responds forming small clusters chain-shaped. If only the static field is present, this is a stable configuration for very long times. The only remarkable trend under these conditions is that the average length of the chains increases with the intensity of the field.

When the perturbation is turned on, a qualitatively different kinetic behavior occurs on the chains leading to a rearrangement of the structure at the long time scale. The perturbation field induces in the chains formed by the static field a waving movement. The amplitude of this movement and the number of nodes generated in the chains depend on the frequency and intensity of the perturbation field and on the length of the chains. These chains movement favor the lateral interactions which induce the aggregation of chains to form larger ones or thicker structures, depending on their relative position. For instance, one observes that if two chains are parallel but shifted along the static field direction, due to the waving the chains extremes with opposite poles may become close enough to attract each other, leading to the  formation of longer chains. 
Another observed situation was that, if two chains are parallel but partially shifted, in such a way that the waving does not approach the poles, but one of them approaches to the inner part of another chain, then, the waving propitiates they aggregate laterally by means of a zippering like motion \cite{furst}. It leads to the formation of longer and partially thicker chains. Finally, if two chains of similar sizes are parallel and they are  one in front of the other, then, the waving induces a lateral aggregation similar, but faster, than that induced by thermal fluctuations. It leads to the formation of thicker structures. 

In order to evaluate the relative importance of the fields in the pattern formation, namely, the effect of the perturbation and that produced by a more intense magnetic field, the distribution of the chain lengths corresponding to several static field intensities were measured and contrasted to those obtained when the dispersion is under the influence of a moderate static field plus a perturbation. The distributions were obtained $200 s$ after the static field was turned on. The measurement of the length of the aggregates were made digitally by means of the software package Sigma Scan Pro 4.0 of Jandel Scientific. All the aggregates smaller than 100 $\mu m$ were ignored.

\begin{figure}[!tbp]
\centering
\begin{center}
\leavevmode 
\psfig{file=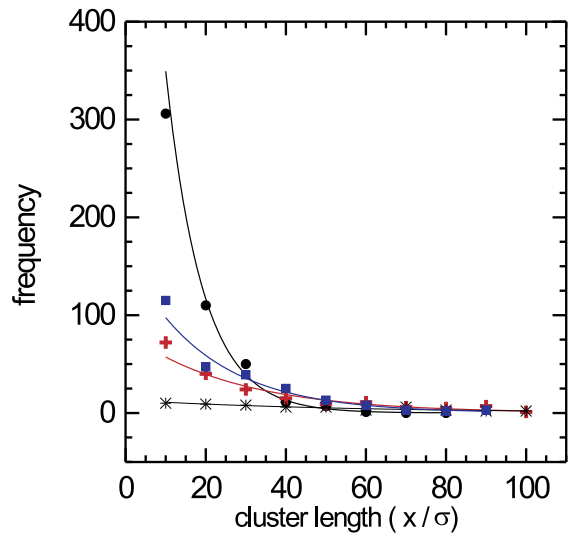,width=7.4cm} \caption{\small Cluster length distribution for some intensities of the static field. Circles, $87 G$; squares, $133 G$; crosses, $157 G$. The asterisks distribution corresponds to a static field of $80 G$ plus a transversal perturbation of $12 G$ $RMS$. The measured length frequencies was partitioned in intervals of width 10$\times \sigma$. The solid lines are exponential fittings.} \label{allinten}
\end{center}
\end{figure}

The cluster length distribution shows an abrupt decreasing behavior, it exhibits a clear dependence on the intensity of the static field. The distributions corresponding to 14 magnetic field strengths were analyzed. In Fig. \ref{allinten} there appears a comparison of the results for $87 G$, $133 G$ and $157 G$. The observed range of the cluster lengths was partitioned in intervals of 10, taking the particle mean size, $\sigma$, as the unit. The symbols in the figure indicate the number of clusters measured within that interval. 
The solid lines are the exponential function that better fits the experimental data respectively. As expected, it is observed  that an increment in the magnetic field intensity leads to a narrower distribution. 

The following expression was used to fit the measurements for the various intensities  of the applied field 

\begin{eqnarray}
R(x/\sigma)=A \exp{\left[-\beta_i (x/\sigma)\right]},
\end{eqnarray}
where $\beta_i$ acquires a different value for each magnetic field intensity and characterizes the sharpness of the distribution. 
 
\begin{figure}[!tbp]
\centering
\begin{center}
\leavevmode 
\psfig{file=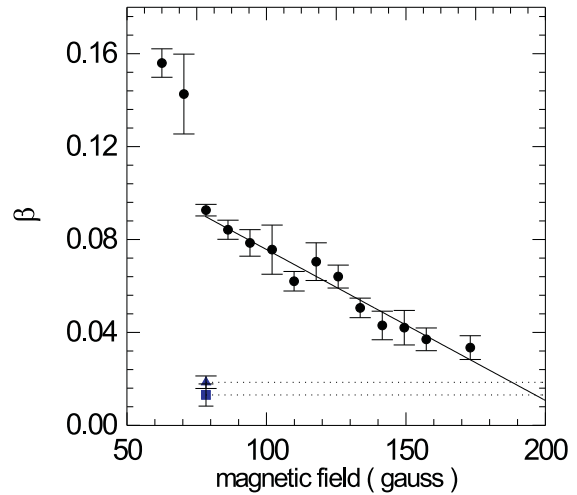,width=7.4cm} \caption{\small The sharpness of the length distribution measured in terms of the exponent $\beta_i$ for various conditions. Circles, various magnetic static fields; triangle, a static field of $80 G$ and after $300 s$ it is added a transversal perturbation field of $12 G$ $RMS$ at $4 Hz$; squares, a static field of $80 G$ plus a perturbation field of $12 G$ $RMS$ at $4 Hz$, simultaneously applied.} \label{expinten}
\end{center}
\end{figure}

Fig. \ref{expinten} shows the exponent  $\beta_i$ as a function of the intensity of the applied field. It is observed that, except for the two first values of $\beta_i$, in a first approximation and in the range of magnetic field here considered, the $\beta_i$ values are directly proportional to the static applied field strength $B_{s}$. 

If both fields are present important differences respect to the previous situation  appear in the structure, both in the short time as well as in the long time scales. For instance, it is observed that the length distribution has exponential characteristics, however, in the presence of both fields it is noticeably narrower, namely, the value of the exponent $\beta_i$ is lower. As an illustration of this, in Fig. \ref{expinten} there appear in the lower left quadrant two values of the exponent $\beta_i$, these correspond to two physical situations when the system is in the presence of a static field of $80 G$ and a perturbation field of $12 G$ $RMS$ with a frequency of $4 Hz$. In the first case, the static field is applied and after $300 s$ the perturbation field is turned on, the corresponding distribution of the cluster length is observed in the Fig. \ref{allinten} (asterisks). In the second case the static field and the perturbation field were applied simultaneously. It was observed that after $200 s$ that both fields begin acting the cluster structure reaches a stable condition where only very slow changes occur. The exponent $\beta$ allows a graphic comparison to appreciate the important differences in the length  distributions. Notice that  the $\beta_i$ values are lower in almost an order of magnitude compared to those values obtained when only the static field  of a similar intensity is applied. Furthermore, these $\beta_i$ values are also smaller than the value reached with highest static field. One may expect that these strong differences in the cluster length distribution when the system is in the presence, or in absence of a perturbation, must have some manifestation in the rheological properties of the dispersion. This will be discussed in a latter section.

\begin{figure}[!tbp]
\centering
\begin{center}
\leavevmode 
\psfig{file=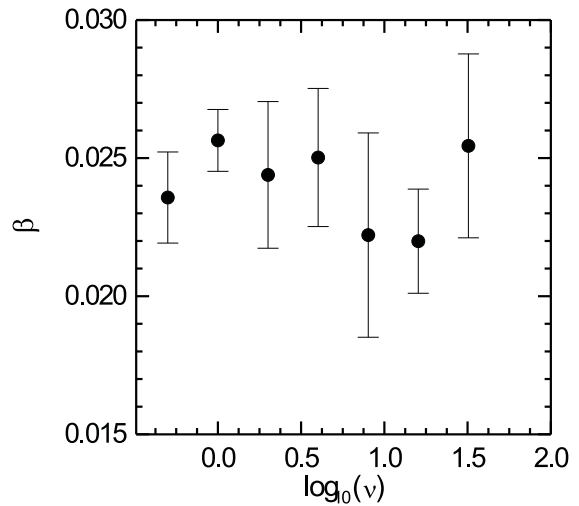,width=7.4cm} \caption{\small The exponent $\beta_i$ as a function of the frequency  of the perturbation, $\nu$. The intensity of the static field and the amplitude of the perturbation remain constant at $80 G$ and $12G$ $RMS$, respectively.} \label{exphz}
\end{center}
\end{figure}

The frequency of the perturbation field has also an interesting role in the determination of the cluster length distribution. In Fig. \ref{exphz} it is depicted the behavior of the exponent $\beta_i$ as a funci\'on of the frequency of the perturbation. In this serie of experiments the perturbation and the static fields were applied simultaneously  and hold fixed at the values $12 G$ $RMS$ at $4 Hz$ and $80 G$, respectively. The distribution was obtained after $200 s$ that both fields begin acting. The frequency of the perturbation was set to the values $0.5, 1, 2, 4, 8, 16$ and $32 Hz$. From the graph it is observed that there is not a simple relationship between the frequency and the exponent $\beta_i$. As it was commented before, the amplitude of the waving motion of the chains is the aspect of the kinetics which propitiates  their aggregation and the building of larger and thicker aggregates. This amplitude depends in a complex way on several factors, one of them is certainly the frequency of the perturbation, but it also must depend on the viscosity of the liquid, the intensity of the perturbation, the particle concentration, and on the length of the chain. At the specific physical conditions presented in Fig. \ref{exphz}, it is found that the narrowest distribution occurs when the perturbation frequency is about 10Hz. 

\begin{figure}[!tbp]
\centering
\begin{center}
\leavevmode 
\psfig{file=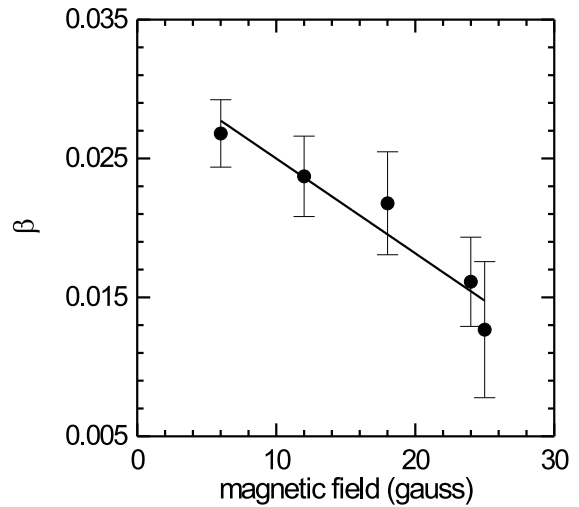,width=7.4cm} \caption{\small The strong influence of the amplitude of the perturbation shown trough the graph of the exponent $\beta_i$ {\em vs}, the amplitude of the perturbation field $B_p$.} \label{expper}
\end{center}
\end{figure}

In Fig. \ref{expper}  it is shown the behavior of the exponent $\beta_i$ as a function of the amplitude of the perturbation field $B_p$. As it was expected, the amplitude of the perturbation has a notorious influence on the value of $\beta_i$. It is observed that $\beta_i$ diminish linearly with the amplitude of the perturbation. By extrapolating the behavior observed in Fig. \ref{expper}, one may conclude that the generation by lateral  aggregation of thicker and larger structures, as it is caused by a  more intense perturbation field, would eventually reach its maximum effect. For the conditions of this sample this maximum effect, or minimum value of $\beta_i$, would be reached at intensities of the order of $40 G$.

It is important to emphasize that this long time behavior of a low concentration MRF has important qualitative and quantitative differences with the behavior observed for higher and moderate concentrations and at short time scales. For these latter conditions the most important general trends found in the pattern formation are the following. The final structure may be seen as composed statistically by clusters of three different generations which are formed sequentially \cite{pre}. At these conditions, a characteristic time of the mechanical response of the system is lower than 0.1 s. On the other hand, in the regime of low particle concentration and long time scale, when a static field is applied the rheological response, or equivalently, the pattern formation, may be described as composed of two sequential stages: in the first one the viscosity, due to the formation of chains, has a relatively rapid increase, this stage could take typically about one minute. During the second stage the patter formation  and the change in the viscosity are basically governed by the lateral aggregation of chains. This slower process could take dozens of minutes or even hours. This response of the MRF can be changed at any stage of its evolution by means the application of an oscillatory  magnetic perturbation. 

As far as now, it has been described the formation of clusters chain-shaped and the influence of the perturbation field on the aggregation processes. Both fields were applied in the horizontal plane of the microscope stage. The glass cell containing the MRF was large enough to minimize the influence of the confinement in the horizontal plane. These observations were done trying to obtain some insight in the kinetics and basic mechanisms by which the cluster structure is built in the suspension leading to changes in the rheological behavior of the system. However, there exist some aspect of this experimental analysis of the rheology of a MRF worthy of some comment. It regards the configuration of the applied fields and the confinement, this later is caused by the small separation between cone and plate of the rheometer, in this case 3 mm. To conduct measurements of viscosity there was selected a configuration of fields that allows to study the dynamical behavior of the shear stress due to the lateral aggregation of chains. The static field was applied in the vertical direction, and the perturbation field was applied in a horizontal direction by means the coil system used for the observations at the optical microscope as it is shown in Fig. \ref{coils}. To simulate in the glass cell the conditions occurring in the rheometer and observe at the microscope the effect of the confinement, two parallel pieces of glass were allocated in our cell 3 mm apart. Under these conditions the static field was applied perpendicularly to these confinement surfaces. What was observed may be described as follows: With only the static field applied, the first stages in the aggregation processes are much like it was observed in the unconfined case. However, once the chains have reached the size of the separation between the walls, there occurs an enhancement of the lateral aggregation of chains. 
It was observed also that to accelerate this process of lateral aggregation, it is necessary  to apply more intense static fields.
After some time the lateral aggregation produces robust compact structures. 

The most important aspect of this, that is discussed and characterized below, is the fact that the whole process is accelerated and enhanced by the application of the perturbation field. 

\section{Viscosity measurements} 

\begin{figure}[!tbp]
\centering
\begin{center}
\leavevmode 
\psfig{file=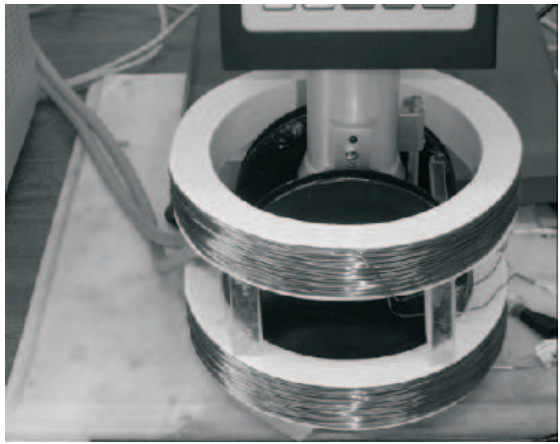,width=7.4cm} \caption{\small Electromagnets arrangement around the column of the cone-plate rheometer. The same coil system is used for the observations at the optical microscope.} \label{coils}
\end{center}
\end{figure}

To measure the viscosity in the MRF samples described in the previous section, a cone-plate rheometer (Brookfield LVDV-III) at $20 ^o C$ (thermal bath Brookfield TC 602P) was used. The experimental set up including the coil system and the rheometer is shown in Fig. \ref{coils}.  

\begin{figure}[!tbp]
\centering
\begin{center}
\leavevmode 
\psfig{file=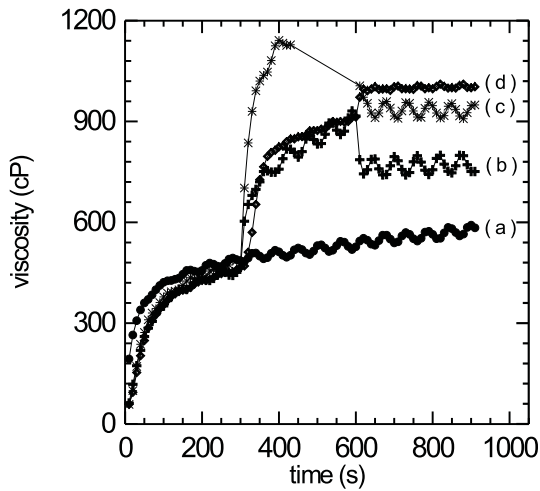,height=6.4cm} \caption{\small Viscosity behavior of a MRF as a function of time for various conditions. (a), only a static field of $80 G$; (b) and (c), a static field of $80 G$ during $300 s$, then it is added with a $92 G$ and $100 G$, respectively, during $300 s$, after that, it is turned back to $80 G$; (d), a static field of $80 G$, at $300 s$ is added with a perturbation field of $12 G$ $RMS$ at $4 Hz$ during $300 s$. } \label{comp1}
\end{center}
\end{figure}

The analysis of the effect starts by comparing the long time rheological behavior of the MRF in the presence and in absence of the perturbation field. Firstly, the viscosity of a dispersion with a concentration $0.05$ in volume fraction is measured when only a static field of  $80 G$ is applied and the rheometer was set at a low cut speed of $1 RPM$. Line (a) of Fig. \ref{comp1} shows the time evolution of the viscosity under these conditions. As expected, there occurs an initial rapid increment in the viscosity, then, an interval of transition to a monotonous increment at a lower rate. This behavior is observed along a time interval of $900 s$. This line exhibits the characteristic fluctuations of a rheological dispersion at a low speed of cut. It will be considered as a reference in the discussion below.
The initial rapid increment can be associated to the stage of the pattern formation driven by the dipolar interaction among the particles. The slower increment in the viscosity is then due to an ulterior reconstruction process induced by the shear and lateral aggregation, leading to larger and thicker clusters in the dispersion.   

In order to observe the role of the intensity of the static field on the viscosity, after some time the intensity is changed. In curves (b) and (c) it is shown the time evolution of the viscosity when initially it is applied a static field of $80 G$, and after $300 s$ this is increased to $92 G$ and $100 G$, respectively.  As it could be expected, what is observed is a second rapid increment in the viscosity starting at $300 s$. The form of the curves at this time interval is qualitatively similar to that occurred at the initial stage with the intensity $80 G$. These $92 G$ and $100 G$ intensities of the static field were held during $300 s$, then, the intensities are tunned back to $80 G$. Under these conditions in (b) and (c) a decrement in the viscosity is measured, however, the viscosity does not reach the values as low as the corresponding ones in the reference curve (a).

To contrast the effect of the magnetic perturbation, firstly it is applied the static field with an intensity of $80 G$ during $300 s$,  then, the transversal  oscillatory perturbation. This later had an intensity of $12 G$ $RMS$ and frequency of $4 Hz$.  The result of these measurements is depicted in curve (d). In a similar way as it was observed  in (b), occurs an increment in the viscosity when the perturbation field is turned on. However, when the perturbation is suppressed, a qualitative different behavior occurs. Unexpectedly, an increment in the viscosity is observed and the values reached by the viscosity remain during the lasting $300 s$ of the measured interval. From these results one concludes that the effect of the perturbation on the viscosity goes beyond the effect of a simply more intense field and that this must be associated to some changes induced by the perturbation in the cluster structure. Since the effect of the perturbation field persists, the cluster formed in  the presence of the perturbation are stable.  

\begin{figure}[!tbp]
\centering
\begin{center}
\leavevmode 
\psfig{file=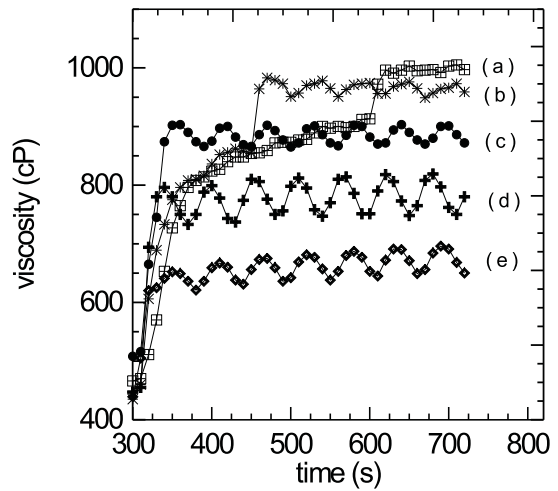,height=6.4cm} \caption{\small Effect on the viscosity of the application of the perturbation during different times. A static field of $80 G$ was turned on and kept during the measurements. At $300 s$ the perturbation field of $12 G$ $RMS$ at $4 Hz$ was applied during (a) $300 s$, (b) $150 s$, (c) $30 s$, (d) $10 s$, and (e) $5 s$.} \label{times}
\end{center}
\end{figure}

A pertinent question in this context is, how long it takes the rheological response of the dispersion under the perturbation field? To answer this question a series of experiments was conducted measuring the viscosity while the dispersion was exposed to a static field of intensity $80 G$ and the perturbation was applied during different intervals of time. Firstly the static field was applied, at the time $300 s$ the perturbation was turned on during $300$, $150$, $30$, $10$, $5 s$,  respectively. In all these cases the frequency of perturbation was $4 Hz$ and the amplitude $12 G$ $RMS$. The results of these measurements are shown in Fig. \ref{times}. Therefore, it is observed that in the present conditions of particle concentration and particle mean size, it takes about $30 s$ the structural changes induced by the perturbation. This time would be shorter for higher particle concentration. 

\begin{figure}[!tbp]
\centering
\begin{center}
\leavevmode 
\psfig{file=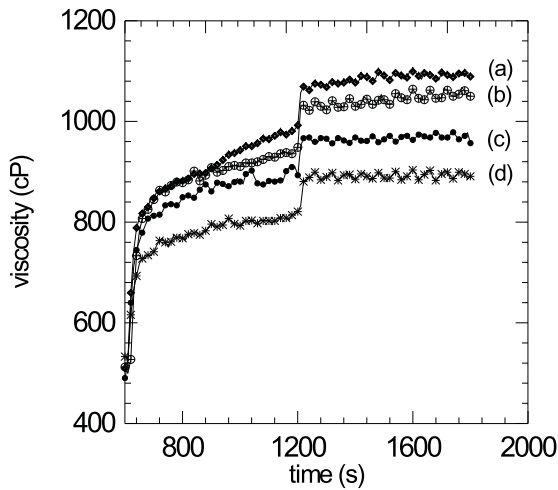,height=6.4cm} \caption{\small Effect of the frequency of the perturbation on the viscosity behavior of a MRF. A static field of $80 G$ was present during the measurements. The perturbation field of $12 G$ $RMS$ was applied at the time $600 s$ and was kept during $600 s$. The frequency of the perturbation was (a) $1 Hz$, (b) $2 Hz$, (c) $4 Hz$, and (d) $16 Hz$. } 
\label{freq}
\end{center}
\end{figure}

Since the waving movement of the chains is the mechanism that more effectively produces the lateral aggregation, it is interesting to obtain some insight about the effects of the frequency. At some extent this analysis could provide some information about the characteristic response time associated to the different chain lengths. A serie of experiments was done in the same sample, applying first a static field of $80 G$. After a time $600 s$ the perturbation field of amplitude $12 G$ $RMS$ was turned on and kept during $600 s$, this was done for various frequencies, $0.5$, $1$, $2$, $4$, $8$, and $16 Hz$. Fig. \ref{freq} shows the behavior of the viscosity as a function of time under these conditions. 

For the conditions of this sample the largest value of the viscosity is reached with the frequency of the perturbation between 1 Hz and 2 Hz. An interesting trend exhibited here is that for higher frequencies, namely, in the interval (2 Hz-16 Hz), the viscosity behaves as a decreasing function of the frequency. Since the rate of the lateral aggregation of chains depends on the amplitude of the waving movement, this likely  indicates that if the mineral oil, in which the chains are embedded were less viscous,  the maximum value of the effective viscosity, reached by the dispersion in the presence of the fields, it would be reached at larger frequencies. 

\begin{figure}[!tbp]
\centering
\begin{center}
\leavevmode 
\psfig{file=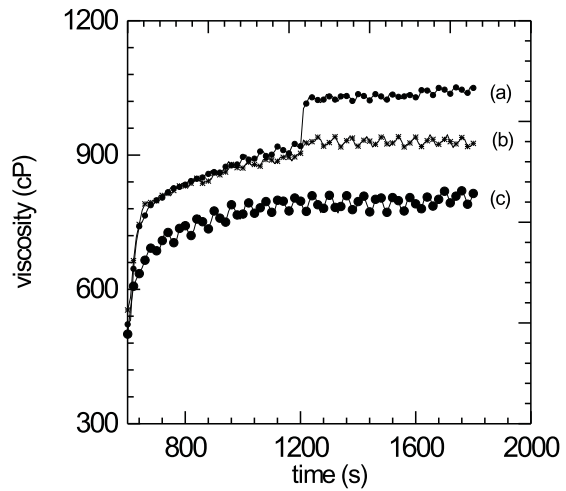,height=6.4cm} \caption{\small The viscosity as a function of time for a system exposed to a static field of $80 G$ and a perturbation of $4 Hz$, applied at the time $600 s$ and kept during $600s$. The amplitude in (a) was $6 G$ $RMS$, in (b) $12 G$ $RMS$, and in (c) $18 G$ $RMS$.} 
\label{amplit}
\end{center}
\end{figure}

\begin{figure}[!tbp]
\centering
\begin{center}
\leavevmode 
\psfig{file=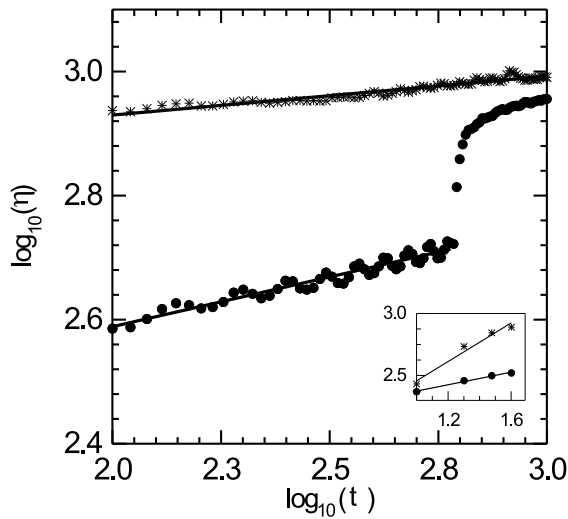,width=7.4cm} 
\caption{\small $Log_{10}-log_{10}$ graph of the viscosity values {\em vs} time. Lower curve, a static field of $80 G$ was applied, then at the time $600 s$ the perturbation field of $12 G$ $RMS$ and $4 Hz$ was turned on. Upper curve, a $80 G$ static field and a $12 G$ $RMS$ with frequency $4 Hz$ perturbation were applied simultaneously. The solid lines are linear fittings. Inset: comparison of the short time measurements of the viscosity.} 
\label{scal}
\end{center}
\end{figure}

It is also interesting to inquire about the role of the amplitude of the perturbation field. Measurements of viscosity as a function of time were done for three different values of the amplitude: $6 G$, $12 G$, and $24 G$, all of them with the same frequency $4 Hz$. Firstly, a $80 G$ static field was applied. At the time $600 s$ the perturbation field was turned on and kept during $600 s$. The results of these measurements are depicted in Fig. \ref{amplit}. The expected changes in the viscosity, namely,  increasing amplitudes lead to increasing values of viscosity, are evident in the curves. In terms of the chains, it is expected that increasing amplitudes of the perturbation  lead to larger and thicker clusters in the dispersion. \\
On the other hand, when the perturbation is turned off merge in the rheological behavior interesting differences between these cases. For the the amplitude (a), after we remove the perturbation field the viscosity measurements remains almost without changes, however for the two higher amplitudes, (b) and (c), we observe an increment in the viscosity. It means that for these amplitudes the perturbation induces changes in the structure, but these changes do not lead to well consolidated clusters. Likely is the waving movement that denies the consolidated structures breaking the chains for certain amplitudes. Then, the structures reach their lowest energetic state until the perturbation is turned off. \\ 
As far as now, it has been analyzed the effect on the viscosity of the application of a static field, followed after some time by a perturbation. However the comparison of these results with that obtained when simultaneously both fields are applied, reveals an interesting new aspect of the rheological behavior of dilute magnetic dispersions. In Fig. \ref{scal}  in a $log_{10} - log_{10}$ graph there appears the long time behavior of the viscosity, for times longer than $100 s$. The lower curve shows the increasing values of the viscosity if at the initial time is only present a static field of $80 G$ and at the time $600 s$ the perturbation field is applied. The upper curve was obtained when both fields were simultaneously applied and kept along the whole interval of time, $0-1000 s$. A static field $80 G$ and an oscillatory perturbation $12 G$ $RMS$ and $4 Hz$ were used. 
In the long time scale, it is clear that a power law describes well the behavior of the viscosity as a function of time, however a remarkable difference between the curves is the value reached by the viscosity at the time $100 s$. Since in both situations the dispersion starts from the same initial conditions, this difference indicates that the increasing rate of the viscosity must be related to the kinetic mechanism of lateral aggregation induced by the perturbation, namely, the waving motion. This can be corroborated in the steep increment of the viscosity shown in the lower curve at the time $600 s$ when the perturbation was applied. The values of the exponents for the corresponding fitting power laws are $0.160$ and $0.062$ for the lower and upper curves, respectively.\\ 
Thus, the short time rheological response is strongly affected by the perturbation. The inset of Fig. \ref{scal} shows a $log_{10} - log_{10}$ graph of the divergent behavior of the viscosity at a time interval $10-60 s$.\\
In the low shear speed regime $0.5-1.5$ $RPM$ and for low concentrated dispersions,  the shear has  only a slight influence on the rheological response of the system. The general trend is that the viscosity reaches relatively higher values for lower cut speed.

\section{Comments and Remarks}
The aim of this work was to characterize the effects of an oscillating field of low intensity on the structure formed by the particles and the rheological behavior  of a  MRF which is under the influence of a static magnetic field. 
The waving movements induced on the chains enhances their lateral aggregation and generates a new kinetic mechanism that change the structure in a way remarkably more intense and faster than that produced by thermal fluctuations and weak magnetic interaction \cite{taostrong,furst,micromech}.
The amplitude and frequency of the oscillating field allow to control at some extent the cluster structure and consequently the rheological behavior of the system. When the perturbation has an amplitude and frequency that does not break the chains, the value reach by the viscosity do not drop after the perturbation field is turned off, as it is observed in Figs. $7-10$. This suggests that the changes induced by the perturbation lead, in these conditions to the formation of structures with a more stable configuration.\\ 
It is worthy of remark the important difference of the short time rheological response shown in Fig. \ref{scal} when simultaneously a static field and a small perturbation are applied, in comparison to that when only a static field is present. This high value of the viscosity can not be reach, in the short time, by a sequential application of the fields. 

 By means of the perturbation it is posible to induce important changes in the cluster structure and on the rheological behavior, still when the system has been long times in the presence of the static field, this can be observed in Fig. \ref{scal}.\\ 
In summary, the results here discussed indicate that in the regimen of low particle concentration, by controlling the amplitude and frequency of the perturbation it is posible to perform a dynamical fine manipulation of the rheological response of a MRF.\\ 
By scaling properly the characteristic of the applied fields, most of the results here discussed would certainly occur in a similar way in MRF composed by larger particles. In principle, whereas thermal fluctuations do not become as important as dipolar interactions, an analogous behavior would also happen in suspensions of shorter particles as well in electro rheological fluids.

\begin{acknowledgments}
On sabbatical leave from Instituto de F\'{\i}sica, Universidad 
Aut\'onoma 
de Puebla, M\'exico. J.L. Carrillo thanks the Secretaria 
de Estado de Universidades e Investigaci\'on of Spain for financial 
support under the grants SAB2005-0063.  J.L.C. thanks Prof. M. 
Rub\'{\i} for hospitality at UB.\\ Partial financial support by CONACyT M\'exico. 
Grant No. 44296 and U.A.P. grant No. 145-04/EXC/G. \\
PROMEP, M\'exico, Grant 103.5/06/1275-234 is acknowledged.
\end{acknowledgments}


\begin{thebibliography}{99}
\bibitem{kitti} D. Kittipoomwong and D. Klingenberg, J. Rheol., {\bf 49},1521 (2005)
\bibitem{pre} J.L. Carrillo, F. Donado, and M.E. Mendoza, Phys. Rev. E {\bf 68}, 061509 (2003)
\bibitem{martin3}J.E. Martin, K.M. Hill, and C.P. Tigges, Phys. Rev. E {\bf 59}, 5676 (1999)
\bibitem{taostrong} R. Tao, J. Phys.: Condens. Matter. {\bf 13}, R979 (2001)
\bibitem{seval} S. Genc, P. Phul\'e, Smart Mater. Struct., {\bf 11},140 (2002)
\bibitem{ref21} P.P. Phule and J.M. Ginder, MRS Bulletin, August, 19, 1998
\bibitem{vicente} J. Vicente, M.T. L\'opez-L\'opez, J.D.G. Duran, and F. Gonzalez-Caballero, Rheol. Acta, {\bf 44}, 94 (2004)
\bibitem{tao} R. Tao and Q. Jiang, Phys. Rev. E {\bf 57}, 5761 (1998)  
\bibitem{chen} S. Chen and C. Wei, Smart Mater. Struct. {\bf 15}, 375 (2006)
\bibitem{martin2} J.E. Martin, J. Odinek, T.C. Halsey, and R. Kamien, Phys. Rev. E {\bf 57}, 756, (1998)
\bibitem{halsey2} T.C Halsey, J.E. Martin, D. Adolf, Phys. Rev. Lett., {\bf 68}, 1519 (1992)
\bibitem{ukai}T. Ukai and T. Maekawa, Phys. Rev. E {\bf 69}, 032501 (2004)
\bibitem{micromech} E.M. Furst and A.P. Gast, Phys. Rev. E., {\bf 61}, 6732, (2000)
\bibitem{furst} E.M. Furst, and A.P. Gast, Phys. Rev. E., {\bf 62}, 6916, (2000)
\bibitem{silva} A.S. Silva, R. Bond, F. Plouraboue, and D. Witz, Phys. Rev. E., {\bf 54}, 5502, (1996)
\bibitem{martin1}J.E. Martin, Phys. Rev. E., {\bf 63}, 011406, (2000)
\bibitem{cutillas2}S. Cutillas and J. Liu, Phys. Rev. E {\bf 64}, 011506 (2001)
\bibitem{stat} J.L. Carrillo, M.E. Mendoza, and F. Donado, J. Stat. Mech. {\bf P06}, P06001, (2005)
\bibitem{dwirtz} D. Wirtz and M. Fermigier, Phys. Rev. Lett., {\bf 72}, 2294, (1994)
\bibitem{joanne} J.H.E. Promislow, and A.P. Gast, Phys. Rev. E., {\bf 56}, 642, (1997)
\bibitem{chaker} M. Chaker, N. Breslin, and J. Liu, Proc. 7th. Int. Conf. on ER Fluids and MR Suspensions, Ed. R Tao (World Scientific, Singapore, 2000). pg. 366.
\bibitem{sonia} S. Melle, G.G. Fuller, and M.A. Rubio, Phys. Rev. E., {\bf 61}, 4111, (2000)
\bibitem{morimoto} Y. Nagaoka, H. Morimoto, and T. Maekawa, Phys. Rev. E., {\bf 71}, 032502, (2005)
\end{thebibliography}
\end{document}